
\input phyzzx.tex
\PHYSREV


\def\O{{\cal O}}
\def\slash#1{\rlap{\hskip .06 em/}#1}
\def\aslash#1{\rlap{\hskip .2 em/}#1}
\def\Ima{{\rm Im\,}}
\def\etal{{\it et. al.}}
\def\pll{|\vec p\,| \leq \Lambda}
\def\pgl{|\vec p\,| > \Lambda}

 \doublespace
 \Pubnum{HU-TFT-93-16\cr hep-ph/9303262}
 \date{March 1993\cr Revised 2 June 1993}

 \titlepage
\vfill
 \title{THE SCHWINGER MODEL AND PERTURBATIVE CHARGE SCREENING}
 \author{Paul Hoyer}
\address{Department of Physics \break 
University of Helsinki, Helsinki, Finland}
\vfill

\abstract

\noindent The well-known exact solution of the massless Schwinger
model can be simply obtained by perturbing around a vacuum without the
Dirac sea of filled negative energy states. The unusual vacuum structure
changes the sign of the $i\epsilon$ prescription at the negative energy
pole of the free fermion propagator. Consequently, all convergent fermion
loop integrals vanish. The logarithmically divergent two-point loop gives
a mass $e/\sqrt\pi$ to the photon, which turns into the free pointlike
boson of the Schwinger model. We discuss the possiblity of using
corresponding expansions to describe charge screening effects in the
massive Schwinger model and the confinement phenomenon of QCD.

\vfill
\endpage

\REF\js{J. Schwinger, Phys. Rev. {\bf 128} (1962) 2425; J. H. Lowenstein
and J. A. Swieca, Ann. Phys. (N.Y.) {\bf 68} (1971) 172; A. Casher, J.
Kogut and L. Susskind, Phys. Rev. {\bf D10} (1974) 732; S. Coleman, R.
Jackiw and L. Susskind, Ann. Phys. (N.Y.) {\bf 93} (1975) 267; S.
Coleman, Ann. Phys. (N.Y.) {\bf 101} (1976) 239; R. Jackiw, {\it
``Topological Investigations of Quantized Gauge Theories"}, in S.
Treiman, \etal, {\it ``Current Algebras and Anomalies"}, World
Scientific, Singapore 1985.}

\chapter{Introduction}

In this paper we show how the well-known exact solution of massless QED$_2$
 -- the Schwinger model\refmark{\js} -- can be simply obtained in
perturbation theory. We consider the perturbative expansion around a
vacuum without a Dirac sea, \ie, with the negative energy states left
unfilled. All fermion loop diagrams vanish, except the one with only two
current insertions. Due to its logarithmic singularity the two-point loop
gives a mass $M=e/\sqrt{\pi}$ to the photon. The screening of electric
charge is complete and local, resulting in the dynamics of a free,
pointlike massive
boson.

The interpretation of this result is quite simple. Since the empty vacuum
has no filled (negative energy) states, there is no real or virtual pair
production. This suppresses the fermion degrees of freedom that are present
in expansions around the Dirac vacuum. The non-vanishing, pointlike
contribution of the two-point loop depends on the regularization of its
logarithmic singularity. We regularize by defining the empty vacuum as the
limit of one where the fermion states are filled only for energies $E
\leq - \Lambda$, with $\Lambda \to \infty.$ The pointlike contribution
of the loop is independent of $\Lambda$, and can be obtained from the
standard, gauge-invariant regularization of the two-point function for
$\Lambda = 0.$

The quantitative success of this approach to the Schwinger model suggests
that charge screening effects can perhaps be analogously described using
perturbation theory also when the screening distance is finite, as in the
massive Schwinger model and in QCD$_4$. For a screening distance of
order $1/\Lambda$, all components of the vacuum state with $\pgl$ should
then be standard, \ie, the positive (negative) energy states with large
momentum should be empty (filled). To preserve charge conjugation
symmetry, the $\pll$ part of the vacuum wave function
can be taken as a superposition of two orthogonal states, one with all
(positive and negative energy) states empty and another one with all these
states filled.

In the Feynman rules, a change in the occupation number of a vacuum state
implies reversing the $i\epsilon$ prescription at the corresponding
(positive or negative energy) pole of the free fermion propagator.
Changing the $i\epsilon$ prescription only for $\pll$ leaves the short
distance behavior, and hence the renormalization properties, of the theory
intact. A similar modification can be applied also to the gluon
propagator, even though we do not explicitly know the gluonic vacuum wave
functional that would correspond to this prescription. We shall
comment on some aspects of the perturbative expansion that results when
such changes in the $i\epsilon$ prescription of the propagators are made.

\REF\fj{
R. Floreanini and R. Jackiw, Phys. Rev. {\bf D37} (1988) 2206.}
\REF\ph{
P. Hoyer, Int. J. Mod. Phys. {\bf A4} (1989) 963 and {\bf A4} (1989)
4535.}

The connection between the filling of the negative energy states in the
perturbative vacuum and the form of the free propagator can be seen in the
Schr\"odinger picture	of the Grassmann path
integral\refmark{\fj,\ph}.  The standard Dirac vacuum, which has all
negative energy states filled, has the wave function
$$V_F(t) = {\rm\exp}\left[-{1\over 2} \int {d^3\vec p\over(2\pi)^3}
{\bar\psi} (t,-{\vec p}\,) {{\vec p}\cdot {\vec\gamma} + m\over E_p} \psi
(t,{\vec p}\,)\right] \eqn\dirac$$
Here ${\bar \psi}$ and $\psi$ are the
standard Grassmann variables describing the fermion fields and $E_p
=\sqrt{\vec p^{\,2}+m^2}$.  The Dirac vacuum implies the standard Feynman
propagator,
$$S_F(p) = i
{{\slash p} + m\over (p^0-E_p+i\epsilon)(p^0+E_p-i\epsilon)}\eqn\sfeyn$$
which describes the propagation of free physical fermions.
Particles of positive energy propagate forward in time, while those of
negative energy propagate backwards, and are interpreted as forward moving
anti-particles of positive energy.

\REF\mccartor{
G. McCartor, Z. Phys. {\bf C36} (1987) 329.}

In a regime like that in the Schwinger model, or in the confinement region
of QCD, the electrons (or quarks) are not, however, simply related to the
physically relevant degrees of freedom.  We may then consider also other
choices of vacua, and propagators, than those given by \dirac\ and \sfeyn.
One can in fact see that the Dirac vacuum functional \dirac\ is not
suitable for the physics of the Schwinger model.  The filling of all
states with $E<0$ leads to a non-local wave function in $x$-space, given by
the Fourier transform of $({\vec p}\cdot{\vec\gamma}+m)/E_p$ in \dirac.
Since in the exact solution of the Schwinger model charge is screened
locally, the long-range correlations of the Dirac vacuum suggest that this
is a poor starting point for a perturbative expansion. Nevertheless, even
though the usual perturbation theory requires summing many diagrams, it
has in fact also been used to
derive some of the Schwinger model results\refmark\mccartor.

When none of the negative energy states are filled the vacuum functional is
$$V_E(t) = {\rm\exp}\left[-{1\over 2} \int {d^3{\vec p}\over
(2\pi)^3}{\bar\psi} (t,-{\vec p}\,)\gamma^0 \psi(t,\vec p\,)\right]
\eqn\empty$$
which is local also in coordinate space.  The corresponding
free fermion propagator  differs from the Feynman propagator \sfeyn\ only
in the $i\epsilon$ prescription at the negative energy pole,
$$S_E(p) = i
{{\slash p} + m\over (p^0-E_p+i\epsilon)(p^0+E_p+i\epsilon)}\eqn\sempty$$
Hence both positive and negative energy fermions propagate only forward in
time, $$S_E(t,\vec p) = \theta(t) {1\over{2E_p}}\left[(\slash p
+m)\exp(-iE_pt) + (\slash p^\dagger -m) \exp(iE_pt)\right] \eqn\setp$$
where $p^0=E_p$. This means that all (non-local)
fermion loops vanish, as the fermion must propagate backwards in
time in some part of the loop. In particular, there are no free
fermion pair production thresholds.
The empty vacuum \empty\ may thus be a good starting point for a
perturbative expansion in situations where asymptotic free fermion states
do not exist, as in the Schwinger model.

In a free fermion vacuum we can choose to fill the negative energy
states, or to leave them empty, separately for each momentum $\vec p$. In
theories such as  QCD, or in QED$_2$ with a finite coupling constant to
mass ratio $e/m$ (the massive Schwinger model), we expect the charge
screening radius to be finite, say ${\cal O}(1/\Lambda)$, where $\Lambda$
is some momentum scale ($\Lambda \simeq\Lambda_{QCD} \simeq 200$ MeV for
QCD).  Then it is natural to use the mixed propagator
$$S_\Lambda(p) =
S_E(p) \theta(\vert{\vec p}\,\vert \le \Lambda) + S_F(p) \theta (\vert{\vec
p}\,\vert > \Lambda) \eqn\smix$$
The ultraviolet behavior of the theory is
thus unaffected by the long-distance charge screening effects, as is
physically reasonable.

In the next section we solve the Schwinger model starting from the mixed
propagator \smix, then taking $\Lambda \to \infty$.  In section 3 we
discuss some properties of the perturbative expansion for massive
QED$_2$ and QCD$_4$ using a mixed propagator like \smix\ with finite
$\Lambda$. Our conclusions are summarized in section 4.

\chapter{The Massless Schwinger Model}

The QED$_2$ partition function is
$$Z = \int {\cal D}(A) {\rm\exp}\left\{i\int d^2x\left[-{1\over 4}
F_{\mu\nu}F^{\mu\nu} -e {-\delta\over
\delta\zeta} {\aslash A}  {\delta\over\delta {\bar\zeta}}\right]\right\}
Z_f[\zeta,{\bar\zeta}]{\Big\vert}_{\zeta={\bar\zeta}=0} \eqn\genf$$
where $Z_f$ is the free fermion generating functional, expressed in terms
of the sources $\zeta,{\bar\zeta}$ of the fermion fields $\bar\psi,\psi$:
$$Z_f[\zeta,{\bar\zeta}] = {\rm\exp} \left[{d^2p\over
(2\pi)^2} {\bar\zeta}(-p) S(p)\zeta(p)\right] \eqn\fermz$$
The $i\epsilon$ prescription of the free propagator $S(p)$ depends on the
wave function of the perturbative vacuum used as a boundary condition at
$t = \pm \infty$\refmark\ph.  Here we wish to show that the physics of the
massless Schwinger model is obtained using the unconventional propagator
\sempty.

\REF\epb{T. Eller, H.-C. Pauli and S. J. Brodsky, Phys. Rev. {\bf D35}
(1987) 1493, and references therein.}

Since  both poles of the propagator \sempty\ are in the lower half $p^0$
plane, the loop momentum integral of any convergent fermion loop vanishes
(the $p^0$ contour can be closed in the upper half plane).  Hence the only
potentially non-vanishing fermion contribution to \genf\ is from
the loop with two photon insertions, which is logarithmically divergent in
1+1 dimensions.  Summing these contributions gives
$$Z = \int{\cal D}(A) {\rm\exp} \Bigl\{i \int d^2 x \Bigr[ -{1\over
4}F_{\mu\nu} F^{\mu\nu} - {i\over 2}  \int d^2 x d^2y A_\mu(x)
L_E^{\mu\nu}(x-y) A_\nu(y)\Bigr]\Bigr\} \eqn\geni$$
where $L_E^{\mu\nu}$ is the two-point fermion loop
$$L_E^{\mu\nu}(x-y) = - (-ie)^2
Tr[\gamma^\mu S_E(x-y)\gamma^\nu S_E (y-x)] \eqn\lemp$$
This loop can give a finite contribution only owing to its logarithmic
divergence in momentum space, and we may thus expect that it is
proportional to $\delta^2(x-y)$. This can be seen more explicitly from the
expression of the massless propagator \setp\ in coordinate space:
$$S_E(x^0,x^1; m=0) = \theta(x^0) [{1\over 2} (\gamma^0 +
\gamma^1)\delta(x^0+x^1)+{1\over 2} (\gamma^0-\gamma^1)\delta(x^0-x^1)]
\eqn\setx$$
In the loop \lemp, $\theta(x^0-y^0) \theta(y^0-x^0)$ forces
$x^0=y^0$, while the $\delta$-functions in the propagator \setx\ then
ensure $x^1=y^1$. Hence the Schwinger model boson, viewed as an $f\bar f$
composite, has a pointlike wave function. This feature is known from
studies of the QED$_2$ bound state wave functions in the $m \to 0$
limit\refmark\epb.

Since the full contribution to the loop \lemp\ comes from the local
point $x=y$, its value (and therefore also the mass of the free boson)
is dependent on the regularization procedure. We shall use the
standard regularization which preserves gauge invariance. We
calculate the loop using the mixed propagator \smix, and take $\Lambda
\to \infty$ at the end. Hence the ultraviolet contribution is always
evaluated with the standard Feynman $i\epsilon$ prescription.
This is equivalent to keeping a finite mass in evaluating the loop.
The screening radius $1/\Lambda$ for massive QED$_2$ can be
estimated as the distance at which the linear potential energy
$V(1/\Lambda) = {1\over 2} e^2/\Lambda$ equals the mass $2m$ of a produced
fermion pair,  $$\Lambda \simeq e^2/4m \eqn\rscreen$$ For our present
purposes the precise value of $\Lambda$ is not important, only the
property $\Lambda \to \infty$ as $m/e \to 0$.

For a finite screening radius $1/\Lambda$ we use the
propagator \smix, and thus consider
$$L_\Lambda^{\mu\nu}(x-y) = - (-ie)^2 Tr [\gamma^\mu
S_\Lambda(x-y)\gamma^\nu S_\Lambda (y-x)] \eqn\llam$$
We first calculate the standard, gauge-invariant expression for the
two-point function $L_\Lambda^{\mu\nu}$   when $\Lambda = 0$, \ie,
using Feynman propagators.  This takes care of the ultraviolet
regularization. The expression for $L_\Lambda^{\mu\nu}$ at any finite
$\Lambda$ is then unambiguous, since the loop momentum integral is
modified only over a finite range $\vert p^1\vert < \Lambda$.

The momentum space expression for  $L_0^{\mu\nu}$ can be written as
$$L_0^{\mu\nu}(q) = e^2 {\int_{-\infty}^\infty} dt{\int_{-\infty}^\infty}
{dk\over 2\pi} Tr \left[\gamma^\mu S_F (t,k_+) \gamma^\nu
S_F(-t,k_-)\right]\exp\left(iq^0t\right) + C^{\mu\nu}  \eqn\lzer$$
where $k_\pm \equiv k\pm q^1/2$. From \sfeyn\ we have
$$S_F(t,p^1)={1\over 2E}\left[ \theta(t)(\slash p +m)e^{-iEt}-
\theta(-t)(\slash p^\dagger -m)e^{iEt}\right] \eqn\sftk$$
where $p^0 = E = \sqrt{(p^1)^2+m^2}$. It is straighforward to check that
the integral in \lzer\ actually is convergent.  Due to the logarithmic
divergence of other integral representations of $L_0^{\mu\nu}$, the
integral in \lzer\ needs to  give the usual gauge-invariant definition of
$L_0^{\mu\nu}$ only up to a $q$-independent constant $C^{\mu\nu}$, as
indicated in \lzer.

Some details of the evaluation of the integral in \lzer\ are given in the
Appendix. The result is
$$L_0^{\mu\nu}(q) ={i\,e^2\over \pi} q^2 \left(-g^{\mu\nu} +
{q^\mu q^\nu \over q^2}\right) {\int_0^1}dx {x(1-x)\over m^2-q^2
x(1-x)-i\epsilon} +{i\,e^2\over\pi} g^{\mu1}g^{\nu1} + C^{\mu\nu}
\eqn\lzexp$$
Hence gauge invariance demands
$$C^{\mu\nu} = -i\,{e^2\over \pi}\, g^{\mu1} g^{\nu1} \eqn\cmn$$
The expression for $L_0^{\mu\nu}$ given by \lzer\ is then the
usual one which is directly obtained in a gauge-invariant
procedure such as dimensional regularization.

The screened loop $L_\Lambda^{\mu\nu} (q)$ is given by \lzer\ with the
Feynman propagators $S_F$ replaced with the screened propagators
$S_\Lambda$ of \smix. Since $S_E(t,p) \propto \theta(t)$ according to
\setp, the $k$-integral of  $L_\Lambda^{\mu\nu}$ is restricted to the
domain where at least one of the propagators is a Feynman propagator.
When both propagators are of the Feynman type, \ie, $\vert k_+\vert >
\Lambda$ and $\vert k_-\vert > \Lambda$, the integral is convergent as
before, and the condition on the integration range forces this part of
the integral to vanish as $\Lambda \to \infty$. The finite integration
range where $\vert k_+\vert > \Lambda$ but $\vert k_-\vert < \Lambda$,
and {\it vice versa}, gives a non-vanishing contribution in the $\Lambda
\to \infty$ limit. As shown in the Appendix, the result is
$${\lim_{\Lambda\to \infty}}
L_\Lambda^{\mu\nu}(q) = -i\,{e^2\over \pi} (- g^{\mu\nu}+{q^\mu q^\nu
\over q^2}) \eqn\linfx$$
Substituting this into the partition function \geni\ we
obtain
$$Z = \int {\cal D} (A){\rm\exp}\left\{i \int d^2x
\left[-{1\over 4}F_{\mu\nu}F^{\mu\nu}-{e^2\over
2\pi}A_\mu(x)\left(-g^{\mu\nu}+
{\partial^\mu\partial^\nu\over \partial^2} \right) A_\nu(x)
\right] \right\} \eqn\parf$$
This agrees with the well-known exact solution of the massless Schwinger
model\refmark\js.

\chapter{Finite Screening Lengths in Perturbation Theory}

The solution of the massless Schwinger model presented above has some
interesting aspects:

\noindent (i) Somewhat paradoxically, it shows that perturbation theory
can give simple and correct results even in a strong coupling $(e/m \to
\infty)$ regime, provided the expansion is made around the proper vacuum
state.

\noindent (ii) The derivation suggests an immediate generalization to
higher dimensions.

\REF\gribov{V. N. Gribov, University of Lund preprint LU TP 91-7
(March 1991).}

Consider, then, a system  with finite screening length,
such as the massive Schwinger model or QCD in $3+1$  dimensions.  The
straightforward generalization of the method employed in Section 2 is to
use perturbation theory with screened propagators of the form \smix, with
a momentum scale $\Lambda$ of order \rscreen\ in QED$_2$ and $\Lambda
\simeq \Lambda_{QCD}$ in QCD$_4$.  Here we wish to make some
preliminary remarks concerning the properties of such an expansion. More
careful studies will show whether the effects of confinement can indeed be
described so simply, using perturbation theory. Related ideas, concerning
changes in the occupation of low momentum positive and negative energy
vacuum states, have been put forward in Ref. \gribov.

\section{Quark and Gluon Vacuum Functionals}

The free fermion vacuum functional which leads to the screened fermion
propagator  \smix\ is according to \dirac\ and \empty\ given by
$V_\Lambda^+(t)$, where
$$\eqalign{V_\Lambda^\pm(t) = \exp\biggl\{-{1\over
2} \int {d^3\vec p\over(2\pi)^3} {\bar\psi} (t,-{\vec p}\,)\biggr.&
\biggl[\theta(\pgl) \,{{\vec p}\cdot {\vec\gamma} + m\over E_p}\biggr.\cr
 &\biggl.\biggl.\pm \theta(\pll)\,\gamma^0 \biggr] \psi (t,{\vec p}\,)
\biggr\}\cr} \eqn\vlamb$$
The vacua $\ket{\Omega_\Lambda^\pm}$ described by the wave functionals
\vlamb\ have their positive and negative energy states for $\pll$ all
empty ($\ket{\Omega_\Lambda^+}$) or filled ($\ket{\Omega_\Lambda^-}$).
Charge symmetry considerations suggest that we choose the linear
superposition $$\ket{\Omega_\Lambda} =
{1\over\sqrt2}\,\left(\ket{\Omega_\Lambda^+}
+\ket{\Omega_\Lambda^-}\right) \eqn\omlamb$$ as the ground state for a
perturbative expansion when the physical screening length is $1/\Lambda$.

The states $\ket{\Omega_\Lambda^+}$ and $\ket{\Omega_\Lambda^-}$ have
different occupation numbers over a finite range of $\vec p$, and are
therefore orthogonal\refmark\fj. This orthogonality ensures that there are
only diagonal contributions to Green functions at any finite order of
perturbation theory, \ie, $$\bra{\Omega_\Lambda}\O\ket{\Omega_\Lambda}=
\half\left[\bra{\Omega_\Lambda^+}\O\ket{\Omega_\Lambda^+}+
\bra{\Omega_\Lambda^-}\O\ket{\Omega_\Lambda^-}\right]\eqn\omop$$
for all operators $\O$. Thus all the $\pll$ propagators in a given
Feynman diagram have the same $+i\epsilon$ or $-i\epsilon$ prescription
at both $(p^0=\pm E_p)$ poles. The full result is then obtained by
averaging the $+i\epsilon$ prescription at both poles of
each propagator (as in \sempty) with the $-i\epsilon$ prescription,
Feynman diagram by Feynman diagram.

It is natural to postulate the same $i\epsilon$
prescription for describing the screening of the color charge of gluons.
Thus, for example, the $\bra{\Omega_\Lambda^+}\O\ket{\Omega_\Lambda^+}$
contribution to \omop\ is obtained with a screened gluon propagator
analogous to \smix,
$$D_\Lambda(p) = D_E(p)\theta(\vert{\vec
p}\,\vert \leq \Lambda) + D_F(p) \theta  (\vert{\vec p}\,\vert > \Lambda)
\eqn\gmix$$ where in Feynman gauge
$$iD_E^{\mu\nu}(p) = {-ig^{\mu\nu} \over
(p^0-E_p+i\epsilon) (p^0+E_p+i\epsilon)} \eqn\gempty$$
and $iD_F(p)$ is the usual Feynman propagator.  We do not explicitly know
what gluon vacuum functional would give the propagator \gmix. A free
gaussian functional leads uniquely to a gluon propagator of the Feynman
form\refmark\ph.  However, the consequences of using screened
gluon propagators \gmix\ can be studied even without knowing the
corresponding structure of the gluon vacuum.

The fact that the propagators \smix\ and \gmix\ differ from the
standard Feynman ones only at low momenta $(\pll)$ suggests that the short
distance structure and renormalizability of the screened theory remain
conventional.  This is  obviously important for both theoretical and
phenomenological reasons. On the other hand, low energy production
thresholds of free quarks and gluons are eliminated, as we shall see below.

\section{Gluon and Quark Masses}

Consider again the two-point fermion loop contribution \llam, now for
QCD$_4$.  The standard result, using Feynman propagators $(\Lambda = 0)$
for SU(N) is
$$L_0^{\mu\nu}(q) = iN\delta^{ab}\left(-g^{\mu\nu} + {q^\mu
q^\nu\over q^2} \right)  q^2\omega_0(q^2) \eqn\omdef$$
$$\omega_0(q^2) = -{g^2\over 2\pi^2} \int_0^1 dx\,
x(1-x){\rm\log}[1-x(1-x)q^2/m^2] \eqn\omz$$
where $a,b$ are the color indices of the external currents. In the
current rest frame $({\vec q} = 0)$, the expression for the loop with
screened propagators \smix\ is
$$\eqalign{ L_\Lambda^{\mu\nu} (q)= L_0^{\mu\nu} +
iN\delta^{ab}g^2\int {d^4k\over(2\pi)^3} &\theta(\Lambda^2-{\vec k}^2)
{Tr[\gamma^\mu({\slash k}_+ + m)\gamma^\nu({\slash k}_- +m)] \over (k_+^0
-E_k + i\epsilon)(k_-^0 -E_k + i\epsilon)} \cr & \times \left[
{\delta(k_+^0 +E_k)\over k_-^0 + E_k} +
        {\delta(k_-^0 +E_k)\over k_+^0 + E_k} \right]\cr} \eqn\ldiffint$$
where $k_\pm = k \pm {1\over 2} q$.  The difference between $L_\Lambda$ and
$L_0$ in \ldiffint\ is due to the $i\epsilon$ prescription of the quark
propagator poles at $k_\pm^0 = - E_k = - \sqrt{{\vec k}^2 + m^2}$ for
$\vert {\vec k}\vert \leq \Lambda$, as given by the integral.
Defining $\omega_\Lambda$ as in \omdef\ we get (for ${\vec q} = 0$)
$$\omega_\Lambda = \omega_0 + {4g^2\over 3\pi^2q^2} {\int_0^\Lambda}
{dk\over E_k}\,{k^2(2k^2+3m^2)\over (q^2-4E_k^2 + i\epsilon)} \eqn\omlint$$

It is straightforward to deduce from \omz\ and \omlint\ that
$$\Ima\omega_\Lambda = \cases{0 & for  $q^2 < 4 (\Lambda^2 +m^2)$\cr
  \Ima\omega_0 & for $q^2 > 4 (\Lambda^2 +m^2)$\cr}\eqn\imoml$$ This
result is a direct consequence of the screened quark propagators \smix.
For $\vert {\vec k}\vert \leq \Lambda$ the poles in the loop energy $k^0$
all lie in the lower half plane, preventing a pinch of the integration
contour and hence also an imaginary part.  From a phenomenological point
of view, the absence of an imaginary part  in the low $q^2$ region
reflects the absence of physical quark degrees of freedom. For large
$q^2$, $\Ima\omega_\Lambda$ is as expected given by standard perturbation
theory.

Just as in the Schwinger model, the screened propagators give rise to a
finite mass for the gauge boson.  After an iteration of the quark loop
contribution, the denominator of the gluon propagator is
$q^2(1+\omega_\Lambda)$.  According to \omlint,  $\omega_\Lambda$ has a
pole at $q^2 =0$, shifting the $q^2 = 0$ pole of the free propagator to
$q^2 = m_g^2$, where $\omega_\Lambda(q^2 = m_g^2) = -1$.  For $\Lambda >>
m$ the second term in \omlint\ dominates, and we obtain approximately, for
small $g^2$,
$$m_g^2 \simeq {g^2\over 3\pi^2}\, \Lambda^2 \eqn\mglue$$

     A non-vanishing gluon mass implies a finite range for the color
interaction, as is appropriate for a screened charge. In a quantitative
calculation we should of course include the contribution of the gluon
loop.  Using a screened gluon propagator \gmix, the analytic properties of
the gluon loop would be similar to those of the quark loop.  In
particular, its imaginary part would again vanish for $q^2 < 4\Lambda^2$.

\REF\njl{Y. Nambu and G. Jona-Lasinio, Phys. Rev. {\bf 122} (1961) 345.}

    In an analogous way, the radiative corrections to the quark propagator
give rise  to a finite quark mass of ${\cal O}(g^2 \Lambda)$, even if the
bare quark mass vanishes. Hence chiral symmetry is broken, and one may
consider the applicability of the  Nambu-Jona-Lasinio method\refmark\njl\
in the present framework.

\section{Boost Non-Invariance}

The $\theta$-functions in the screened quark and gluon propagators \smix,
\gmix\  depend on the absolute value of the 3-momentum and hence on the
reference frame. It is thus clear that the Green functions will not be
boost invariant order by order in perturbation theory.

   The non-Lorentz covariance of perturbation theory derives from our
choice of vacuum functional.  For $\pgl$ the
vacuum \omlamb\ includes the usual Dirac sea of negative energy fermion
pairs, while for $\pll$ both the positive and negative energy states are
either empty $(\ket{\Omega_\Lambda^+})$ or filled
$(\ket{\Omega_\Lambda^-})$.  Insofar as the full perturbative  expansion
at least formally represents the complete, Lorentz-invariant theory, one
may hope that boost invariance is restored at higher orders.

\REF\ng{S. J. Brodsky, in Brandeis Lectures 1969, Vol. 1, p. 95 (edited
by M. Chr\'etien and E. Lipworth, Gordon and Breach, N.Y. (1971)); W.
Dittrich, Phys. Rev. {\bf D1} (1970) 3345; A. R. Neghabian and W.
Gl\"ockle, Can. J. Phys. {\bf 61} (1983) 85.}

A well-known example of how Lorentz invariance requires summing
perturbative diagrams to arbitrary order is provided by the Bethe-Salpeter
equation for a light fermion bound to a particle with large mass
$M$\refmark\ng.  The bound state equation has non-invariant projection
operators on the positive energy states.  The relativistically invariant
Dirac equation is obtained in the $M \to \infty$ limit only after the
inclusion of  kernels of arbitrarily high order, thus allowing any number
of fermion pairs in the Fock states. It is also interesting to note that
the Dirac equation is, on the other hand, obtained directly from the lowest
order, single photon exchange kernel when one expands around the empty
vacuum \empty\ (which from the standard point of view already contains an
infinite number of pairs)\refmark\ph.

The dynamics of bound states is very dependent on the choice of frame.
The description of hard scattering processes in terms of perturbative QCD
is generally formulated in a frame where the hadrons have large momenta.
It is only in this frame that their constituents can be treated as
quasifree  partons carrying a measurable fraction of the total momentum.
In the formulation of perturbation theory discussed above, high momentum
$(\vert{\vec p}\,\vert > \Lambda)$ quarks and gluons are treated as
physical particles, obeying the rules of ordinary perturbation theory.
Slow partons -- the constituents of hadrons at rest or wee partons of fast
hadrons -- are ``screened'', in analogy to the  fermions of the Schwinger
model.  The consequent loss of explicit boost invariance is, at least
superficially, in accord with conventional wisdom.

\section{A Pseudothreshold Singularity}

The fermion loop $L_\Lambda^{\mu\nu}(q)$ in
\llam\ has both a threshold singularity in the region
$q^2>0$, and a pseudothreshold singularity when $q^2<0$.
The latter exists in a frame with center-of-mass motion $(\vec q\neq 0)$
when one of the propagators in the loop has momentum above $\Lambda$, and
the other one momentum below $\Lambda$. There can then be a pinch between
the negative energy pole of the Feynman propagator and the negative energy
pole of the $\pll$ propagator \sempty, since they have opposite $i\epsilon$
prescriptions. This corresponds to the excitation of a fermion in the
filled Dirac sea to one of the empty negative energy states. The
pseudothreshold singularity is relevant, \eg, in deep inelastic scattering
(since $q^2<0$), and describes a final state consisting of a ``physical''
antifermion with $\pgl$ and a wee, negative energy fermion with $\pll$.
The charge symmetric process with a fast fermion is obtained from the
$\ket{\Omega_\Lambda^-}$ vacuum component in \omlamb. Some momentum
transfer from a target particle is of course required to materialize the
jet (which starts off with $q^2<0$).

\chapter{Summary}

In this paper we have solved the Schwinger model in the strong coupling
limit $(e/m \to \infty)$, based on perturbation  theory around a vacuum
without the Dirac sea. In such a vacuum fermion pair production is
suppressed, and the sole non-vanishing contribution is given by the
two-point fermion loop, due to its logarithmic singularity. This
expansion is thus much more useful for understanding the physics of the
Schwinger model than the equivalent standard perturbation
theory\refmark\mccartor.

The different vacuum structure manifests itself only in the $i\epsilon$
prescription at the negative energy pole of the free fermion propagator.
Hence the method can be generalized in a straightforward way both to
higher dimensions and to boson propagators.  We briefly discussed some
aspects of such a method for treating the physics of systems with a finite
charge screening length, such as the massive Schwinger model and QCD.
Several interesting features emerge, including finite gluon and quark
masses, which signal the breakdown of gauge and chiral symmetry.  More
work will be required to establish whether such an approach can lead to a
self-consistent and useful description of charge screening by perturbative
methods.

{\bf Acknowledgement:} I am grateful to S. J. Brodsky, H. Hansson, C. S.
Lam and H. B. Nielsen for helpful discussions. I particularly thank J.
Grundberg for pointing out an error in the manuscript.

\appendix

Here we would like to give the (straightforward) derivation of Eqs.
\lzexp\ and \linfx. Substituting the Feynman propagator \sftk\ into \lzer\
and doing the $t$-integral gives
$$\eqalign{ L_0^{\mu\nu} (q) = -{ie^2\over 8\pi}
\int_{-\infty}^\infty {dk\over E_+E_-} &\left\{{Tr[\gamma^\mu({\slash k}_+
+ m)\gamma^\nu({\slash k}_-^\dagger -m)]\over
q^0-E_+-E_-+i\epsilon}\right.\cr
&-\left.{Tr[\gamma^\mu({\slash k}_+^\dagger - m)\gamma^\nu
({\slash k}_- + m)]\over q^0+E_+ + E_- -i\epsilon}\right\}
+ C^{\mu\nu} \cr}\eqn\aone$$
where $k_\pm^0 = E_\pm = \sqrt{k_\pm^2 +
m^2}$, with $k_\pm = k \pm q^1/2$. The traces are ${\cal O}(1)$ for $k \to
\pm \infty$, thus ensuring the convergence of the $k$-integral.

   We express the energies and momenta appearing in \aone\ in terms of a
``center-of-mass'' momentum $p$ as follows:
$$\eqalign{E_\pm &= E_p\cosh\zeta \pm p \sinh\zeta\cr
\pm k_\pm &= E_p\sinh\zeta \pm p \cosh\zeta\cr} \eqn\atwo$$
where $E_p = \sqrt{p^2+m^2}$ and $\sinh\zeta = q^1/2E_p$.  Denoting
the traces in \aone\ by $Tr_1^{\mu\nu}$ and $Tr_2^{\mu\nu}$,
respectively, we find
$$\eqalign{ &Tr_1^{00} = Tr_2^{00} = 4m^2 \sinh^2\zeta\cr
&Tr_1^{11} = Tr_2^{11} = 4m^2 \cosh^2\zeta\cr
&Tr_1^{01} =-Tr_2^{01} = 4m^2 \sinh\zeta \cosh\zeta\cr} \eqn\athree$$
Finally, we choose as the new integration variable $x = (E_p + p)/2E_p$.
Inserting \atwo, \athree\ and the jacobian
$${1\over E_+E_-} \ {dk\over dx} = {2E_p \over m^2\cosh\zeta} \eqn\afour$$
into \aone\ we get \lzexp.

To get \linfx, we start from the expression for the screened loop
$L_\Lambda^{\mu\nu}$ of \llam\ given by \lzer, with the Feynman
propagators $S_F$ replaced with the mixed propagators $S_\Lambda$ of
\smix. The integration range can be divided into four domains, according
to $|k_\pm|$ being $>\Lambda$ or $<\Lambda$. When $|k_+|<\Lambda$ and
$|k_-|<\Lambda$ both propagators are of the $S_E$ type \setp, and the
integral vanishes due to $\theta(t)\theta(-t)=0$. For $|k_+|>\Lambda$ and
$|k_-|>\Lambda$ both propagators are $S_F$, and the integrand is as in
\aone. Since the traces given by \athree\ do not grow with $|k|$, the
integral is convergent and vanishes in the limit $\Lambda \to \infty$.
Hence
$$\eqalign{L_\Lambda^{\mu\nu}(q) = &e^2\int_{-\infty}^\infty dt \int_
{-\infty}^\infty {dk\over 2\pi} \exp(iq^0 t)\Bigl\{\cr
&\theta(|k_+|>\Lambda)\theta(|k_-|<\Lambda)Tr\left[\gamma^\mu
S_F(t,k_+) \gamma^\nu S_E(-t,k_-)\right] \cr
+&\theta(|k_+|<\Lambda)\theta(|k_-|>\Lambda)Tr\left[\gamma^\mu
S_E(t,k_+) \gamma^\nu S_F(-t,k_-)\right]\Bigr\}\cr
+&C^{\mu\nu} + {\cal O}(1/\Lambda^2)}\eqn\llapp$$
where $C^{\mu\nu}$ is given by \cmn. Taking $q^1>0$ and substituting
\setp\ and \sftk\ into \llapp, the first term on the r.h.s. becomes,
$$\eqalign{L_{\Lambda a}^{\mu\nu}(q) = i{e^2\over 8\pi}
\int_{\Lambda-q^1/2}^{\Lambda+q^1/2} {dk\over E_+E_-} &\Biggl\{
{Tr\left[\gamma^\mu(\slash k_+^\dagger-m) \gamma^\nu (\slash k_-+m)\right]
\over q^0+E_++E_- -i\epsilon} \cr
&+{Tr\left[\gamma^\mu(\slash k_+^\dagger-m) \gamma^\nu (\slash
k_-^\dagger-m)\right] \over q^0+E_+-E_- -i\epsilon}\Biggr\}}\eqn\llappa$$
Since the integral is over a finite range in $k$, it can be non-vanishing
in the limit $\Lambda \to \infty$ only if the integrand is finite. The
first term in the integrand of \llappa\ behaves like $1/k^3$ for large
$k$, and hence contributes ${\cal O}(1/\Lambda^2)$ to the integral. Since
$$E_+-E_-=\sqrt{(k+\half q^1)^2+m^2}- \sqrt{(k-\half q^1)^2+m^2} = q^1 +
{\cal O}(1/k^2) \eqn\ediff$$
we get
$$\eqalign{L_{\Lambda a}^{\mu\nu}(q) = i{e^2\over 8\pi}
\int_{\Lambda-q^1/2}^{\Lambda+q^1/2} dk
{Tr\left[\gamma^\mu(\gamma^0+\gamma^1) \gamma^\nu (\gamma^0+\gamma^1)
\right] \over q^0+q^1 -i\epsilon} +&{\cal O}(1/\Lambda^2) \cr
= i{e^2\over 2\pi} {q^1 \over q^0+q^1 -i\epsilon} D^{\mu\nu}+
&{\cal O}(1/\Lambda^2) }\eqn\llappb$$
where $D^{00}=D^{11}=-D^{10}=-D^{01}=1$. The second term in \llapp\
similarly gives (for all $\mu,\nu$)
$$L_{\Lambda b}^{\mu\nu}(q) =-i{e^2\over 2\pi} {q^1\over q^0-q^1
+i\epsilon} + {\cal O}(1/\Lambda^2) \eqn\llappc$$
Summing all contributions in \llapp\ gives the Lorentz and gauge
invariant result
$$\eqalign{L_{\Lambda}^{\mu\nu}(q) &= L_{\Lambda a}^{\mu\nu}(q)+ L_{\Lambda
b}^{\mu\nu}(q) + C^{\mu\nu} + {\cal O}(1/\Lambda^2) \cr
&= -i {e^2\over \pi}(-g^{\mu\nu}+{q^\mu q^\nu \over q^2}) + {\cal
O}(1/\Lambda^2)}\eqn\llappd$$
which is \linfx.
\refout
\end